\documentclass[twocolumn,amsmath,floatfix,prb,aps]{revtex4}

\usepackage{color}
\usepackage{amsmath}
\usepackage{pifont}   % Ding symbols
\usepackage{graphicx} % Include figure files
\usepackage{dcolumn}  % Align table columns on decimal point
\usepackage{bm}       % bold math
\usepackage{amsfonts} % Some more fonts
\usepackage{amssymb}  % More symbols
\usepackage{multirow} % Table functions
\usepackage{romannum}
\usepackage{siunitx}

\usepackage[english]{babel}
\usepackage[autostyle, english = american]{csquotes}
\MakeOuterQuote{"}

\begin{document}
%\DeclareGraphicsRule{*}{png}{*}{}

%%%% User-defined commands %%%%
\newcommand{\ba}{{\bf a}}
\newcommand{\BB}{{\bf b}}
\newcommand{\bd}{{\bf d}}
\newcommand{\br}{{\bf r}}
\newcommand{\bp}{{\bf p}}
\newcommand{\bk}{{\bf k}}
\newcommand{\bg}{{\bf g}}
\newcommand{\bmm}{{\bf m}}
\newcommand{\bj}{{\bf j}}
\newcommand{\bt}{{\bf t}}
\newcommand{\bv}{{\bf v}}
\newcommand{\bu}{{\bf u}}
\newcommand{\bq}{{\bf q}}
\newcommand{\bG}{{\bf G}}
\newcommand{\bP}{{\bf P}}
\newcommand{\bJ}{{\bf J}}
\newcommand{\bK}{{\bf K}}
\newcommand{\bL}{{\bf L}}
\newcommand{\bR}{{\bf R}}
\newcommand{\bS}{{\bf S}}
\newcommand{\bT}{{\bf T}}
\newcommand{\bQ}{{\bf Q}}
\newcommand{\bA}{{\bf A}}
\newcommand{\bH}{{\bf H}}

\newcommand{\bra}[1]{\left\langle #1 \right |}
\newcommand{\ket}[1]{\left| #1 \right\rangle}
\newcommand{\braket}[2]{\left\langle #1 | #2 \right\rangle}
\newcommand{\mel}[3]{\left\langle #1 \left| #2 \right| #3 \right\rangle}

\newcommand{\bdel}{\boldsymbol{\delta}}
\newcommand{\bsig}{\boldsymbol{\sigma}}
\newcommand{\beps}{\boldsymbol{\epsilon}}
\newcommand{\bnu}{\boldsymbol{\nu}}
\newcommand{\bnab}{\boldsymbol{\nabla}}
\newcommand{\bGam}{\boldsymbol{\Gamma}}

\newcommand{\bgt}{\tilde{\bf g}}

\newcommand{\brh}{\hat{\bf r}}
\newcommand{\bph}{\hat{\bf p}}

\title{Spin texture in weak topological insulators: the role of bulk states and band bending}

\author{F. Rost$^1$}
\author{R. Gupta$^2$}
\author{S. Sharma$^3$}
\author{O. Pankratov$^1$}
\author{S. Shallcross$^3$}
\affiliation{1 Lehrstuhl f\"ur Theoretische Festk\"orperphysik, Staudtstr. 7-B2, 91058 Erlangen, Germany}
\affiliation{2 H. H. Wills Physics Laboratory, University of Bristol,Tyndall Avenue, Bristol BS8 1TL, United Kingdom}
\affiliation{3 Max-Born-Institute for Non-linear Optics and Short Pulse Spectroscopy, Max-Born Strasse 2A, 12489 Berlin, Germany}

\date{\today}

\begin{abstract}
We derive the spin texture of a weak topological insulator via a supersymmetric approach that includes the roles of the bulk gap edge states and surface band bending. We find the spin texture can take one of four forms: (i) helical, (ii) hyperbolic, (iii) hedgehog, with spins normal to the Dirac-Weyl cone of the surface state, and (iv) hyperbolic hedgehog. Band bending determines the winding number in the case of a helical texture, and for all textures can be used to tune the spin texture polarization to zero. For the weak topological insulator SnTe, we show that inclusion of band bending is crucial to obtain the correct texture winding number for the (111) surface facet $\Gamma$-point Dirac-Weyl cone. We argue that hedgehogs will be found only in low symmetry situations.
\end{abstract}

\maketitle

\section{Introduction}

One of the most fascinating features of 3d topological insulators (TI) is the relation between the microscopic physics of the bulk and the spin structure of the surface state\cite{fu09,zha12,zha13,liu13,wan13,woj13,shi14}. How bulk electronic structure determines surface spin texture and what texture forms are possible are questions that have been addressed in a number of theoretical treatments\cite{joz11,sou11,moo11,henk12,hai13,pan13,jak15}. These are often based on a continuum description of the bulk, which then requires careful consideration of the boundary condition from material to vacuum\cite{zha12}. The purpose of the present work is to describe a "minimal approach" that treats the important physics of gap inversion and band bending on an equal footing and, furthermore, is analytically solvable. This allows one to trace the microscopic degrees of freedom of the bulk gap edge states into the surface spin texture, and thus to probe the roles of the bulk electronic structure and surface band bending in the formation of the surface state.

To this end we will adopt the supersymmmetry theory of IV-VI semi-conductor heterojunctions developed by Volkov and Pankratov\cite{vol78,vol83,vol85,pan87,pan87a,pan89}. In this approach the bulk is described by a low energy Dirac equation with gap $\Delta_0 < 0$ and the vacuum treated as a large positive gap. For the microscopic basis of the bulk Dirac equation we will consider two cases: (i) gap edge states obtained from a realistic tight-binding model of the IV-VI semi-conductor SnTe, and (ii) gap edge states that are a general expansion in spin and angular momentum

\begin{equation}
\ket{\phi_i} = \sum_{m\sigma} d_{m\sigma}^{(i)} \ket{m\sigma}
\label{1}
\end{equation}
with which we investigate the various spin textures that are, in principle, possible within this model.

We find that band bending plays a key role in the creation of spin texture, as it determines the relative weight of the bulk conduction and valence bands in the surface state. Downward (or upward) band bending leads to dominance of the valence (or conduction) bands in the spin texture and, as a consequence, upon variation of band bending through the bulk gap the winding number of a helical texture can be tuned from +1 to -1 through a texture depolarization point. From the general form for the bulk band edge states, Eq.~\eqref{1}, we show that within a supersymmetric treatment of the TI surface state 4 spin textures are possible: (i) helical, (ii) hyperbolic, (iii) helical hedgehog, and (iv) hyperbolic hedgehog. The hedgehog textures are unusual, and we show that their existence requires a low symmetry of the bulk wavefunction generally not found in topological insulators.

\section{Dirac model of the interface}

We model the material vacuum interface as a Dirac equation
\begin{equation}
H = \begin{pmatrix}
\varphi+\Delta & v_\parallel \sigma_z p_z + v_\perp \sigma_\perp p_\perp \\ v_\parallel \sigma_z p_z + v_\perp \sigma_\perp p_\perp & \varphi-\Delta
\label{H} 
\end{pmatrix}
\end{equation}
where $\bsig\perp = (\sigma_x,\sigma_y)$ and $\bp_\perp = (p_x,p_y)$. The velocities $v_\parallel$ and $v_\perp$ allow for an anisotropic effective masses in the bulk spectrum.

The interface is defined by two $z$ dependent fields ($z$ is the direction of the surface normal), a gap inversion field

\begin{equation}
\Delta = \Delta_0 f(z)
\end{equation}
where the bulk gap $\Delta_0$ is taken to be negative, and a band bending field
\begin{equation}
\varphi = \varphi_0 f(z).
\end{equation}
In these expressions $f(z) \to -F$ as $z \to \infty$ with $F$ a large negative number defining the positive "vacuum gap" $\Delta_0 F$, and $f(z) \to 1$ as $z \to -\infty$. We thus have band inversion at the material-vacuum interface, with the bulk gap $\Delta_0 < 0$ and the vacuum gap $\Delta_0 F > 0$.
Note that it is necessary to take the same $f(z)$ for both $\Delta$ and $\varphi$, justified in the original context of gap inversion semi-conductor hetero-junctions by a common dependence on alloy composition, for example as in the original band-inverting system Pb$_{1-x}$Sn$_x$Te studied by Volkov and Pankratov\cite{vol85}. As the surface state will turn out to be independent of $f(z)$, the assumed common form does not represent a serious approximation in the model.

\section{Derivation of the surface state}

Following Refs.~\onlinecite{vol85,pan87,pan89} we first square the interface equation (Eq.~\ref{H})
generating the expression

\begin{eqnarray}
& & \Bigg[ \Delta^2-\epsilon^2-\varphi^2+2\epsilon\varphi
+v_\parallel^2p_z^2+v_\perp^2p_\perp^2 \nonumber \\
& & + v_\parallel (p_z f(z))\begin{pmatrix} 0 & \varphi_0-\Delta_0 \\ \varphi_0+\Delta_0 & 0 \end{pmatrix} \otimes \sigma_z \Bigg] \Psi = 0
\end{eqnarray}
in which only the last term is non-diagonal. This can be diagonalized by the transformation $S$

\begin{equation}
S^{-1} \begin{pmatrix} 0 & \varphi_0-\Delta_0 \\ \varphi_0+\Delta_0 & 0 \end{pmatrix}  S = \tau_z \sqrt{\varphi_0^2-\Delta_0^2}
\end{equation}
given by

\begin{equation}
S = \frac{1}{\sqrt{2}}\begin{pmatrix} s_- & s_- \\ s_+ & -s_+ \end{pmatrix}
\end{equation}
with

\begin{equation}
s_\pm = \frac{\sqrt{\varphi_0\pm\Delta_0}}{\sqrt{\left|\Delta_0\right|}}
\end{equation}
Introducing the so-called superpotential

\begin{equation}
W(z) = \sqrt{\Delta_0^2-\varphi_0^2}\left(f(z) + \frac{\epsilon\varphi_0}{\Delta_0^2-\varphi_0^2}\right)
\label{S}
\end{equation}
then allows for the factorization of the squared interface equation into the supersymmetric form

\begin{eqnarray}
&&\left(W(z) + i v_\parallel p_z \tau_z\otimes\sigma_z \right)
\left(W(z) - i v_\parallel p_z \tau_z\otimes\sigma_z \right)
\Psi \nonumber \\
& = & \left(\frac{\Delta_0^2\epsilon^2}{\Delta_0^2-\varphi_0^2} - v_\perp^2 p_\perp^2\right)\Psi
\label{F}
\end{eqnarray}

We now demand zero mode solutions of Eq.~\ref{F} by action of the annihilation operator:
\begin{equation}
\left(W(z) - i v_\parallel p_z \tau_z\otimes\sigma_z \right)
\ket{\pm} = 0
\end{equation}
which then gives two possible solutions of Eq.~\ref{F}

\begin{figure*}
\centering
\includegraphics[width=0.95\textwidth]{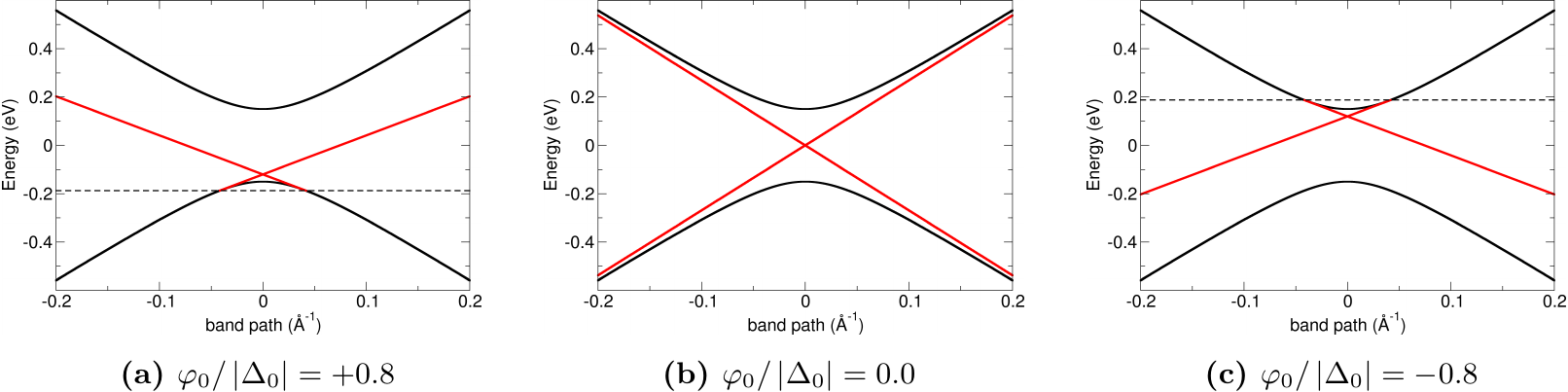}
\caption{Dirac-Weyl surface state for downward, no, and upward band bending. For a finite band bending the Dirac-Weyl surface state smoothly joins the bulk spectrum, while for zero band bending the Dirac-Weyl cone only asymptotically joins with the bulk.}
\label{fig1}
\end{figure*}

\begin{eqnarray}
\ket{+} & = & \ket{\uparrow}\otimes\ket{\uparrow} e^{\frac{1}{\hbar v_\parallel} \int^z\,dz' W(z')} \label{int1} \\
\ket{-} & = & \ket{\downarrow}\otimes\ket{\downarrow} e^{\frac{1}{\hbar v_\parallel} \int^z\,dz' W(z')} \label{int2}
\end{eqnarray}
The zero mode solution of the squared interface equation implies the following condition

\begin{equation}
\epsilon = \pm\sqrt{1-\frac{\varphi_0^2}{\Delta_0^2}} v_\perp p_\perp
\end{equation}
on the interface equation eigenvalues, which we recognize as the Dirac-Weyl spectrum with a velocity reduction factor governed by the ratio of band bending to the bulk gap. For the solutions Eq.~\eqref{int1} and \eqref{int2} to be normalizable requires that the superpotential be real valued and change sign asymptotically. This is guaranteed by the gap inversion at the interface provided the energy $\epsilon > \epsilon_c$ in the case of downward band bending, and $\epsilon < \epsilon_c$ in the case of upward band bending, with $\epsilon_c$ the energy at which the Dirac-Weyl spectrum joins smoothly with the bulk manifold and ceases to exist

\begin{equation}
\epsilon_c = -\frac{\Delta_0^2-\varphi_0^2}{\varphi_0}
\end{equation}
This is illustrated in Fig.~\ref{fig1} for the case of downward (panel (a)), no (panel (b)), and upward (panel (c)) band bending.

As the solutions $\ket{\pm}$ exist in a degenerate sub-space of the squared interface equation $(H^2-\epsilon^2)\ket{\pm} = 0$ we require the coefficients $c_\pm$ that form the linear combination $c_+\ket{+} +c_-\ket{-}$ that solve $H$, which can be obtained by diagonalizing $H$ in the sub-space spanned by $\ket{\pm}$. To this end we first apply the $S$ transformation to $H$
\begin{equation}
S^{-1} H S = \begin{pmatrix}
\varphi + \frac{\varphi_0}{\sqrt{\varphi_0^2-\Delta_0^2}} D
&
\Delta - \frac{\Delta_0}{\sqrt{\varphi_0^2-\Delta_0^2}} D \\
\Delta + \frac{\Delta_0}{\sqrt{\varphi_0^2-\Delta_0^2}} D &
\varphi - \frac{\varphi_0}{\sqrt{\varphi_0^2-\Delta_0^2}} D
\end{pmatrix}
\end{equation}
where $D = v_\parallel \sigma_z p_z + v_\perp \bsig_\perp \bp_\perp$ and then construct the effective Hamiltonian $H_{eff}$ from the functions $\ket{\pm}$

\begin{eqnarray}
H_{eff} = \begin{pmatrix}
\mel{+}{S^{-1}HS}{+} &
\mel{+}{S^{-1}HS}{-} \\
\mel{-}{S^{-1}HS}{+} &
\mel{-}{S^{-1}HS}{-}
\end{pmatrix} 
\end{eqnarray}
This gives us the Dirac-Weyl equation of the surface state

\begin{equation}
H_{eff} = \begin{pmatrix}
0 & i\gamma v_\perp p_- \\
-i\gamma v_\perp p_+ & 0
\end{pmatrix}
\end{equation}
where $p_\pm = p_x \pm i p_y$ and 
$\gamma = \sqrt{1-\frac{\varphi_0^2}{\Delta_0^2}}$
the velocity reduction factor. The coefficients $c_\pm$ are then given by the components of the Dirac-Weyl eigenfunction

\begin{equation}
\begin{pmatrix} c_+ \\ c_- \end{pmatrix} =
\frac{1}{\sqrt{2}} \begin{pmatrix} e^{-i\left(\theta_\bk/2-\pi/4\right)} \\ l e^{+i\left(\theta_\bk/2-\pi/4\right)} \end{pmatrix} e^{i\bk.\br}
\end{equation}
with $l=\pm1$ labeling the electron and hole cones of the spectrum and $\theta_\bk = \tan^{-1} k_y/k_x$. The solution to the interface Hamiltonian $S^{-1} H S$ is then

\begin{equation}
 \Psi=\frac{1}{\sqrt{2}}\begin{pmatrix}
       e^{-i\left(\theta_\bk/2-\pi/4\right)} \\ 0 \\ 0 \\ l                 e^{i\left(\theta_\bk/2-\pi/4\right)}
      \end{pmatrix}
      e^{i\bk.\br}\phi(z)
\end{equation}
which by back transformation with $S$ finally gives us the surface state solution of the original interface equation $H$
\begin{equation}
 \Psi = \Biggl[\frac{c_+}{2}\begin{pmatrix}
                            s_- \\ 0 \\ s_+  \\ 0
                            \end{pmatrix}
              +\frac{l                                             c_-}{2}\begin{pmatrix}
                            0 \\ s_- \\ 0 \\ -s_+ 
                           \end{pmatrix}\Biggr] e^{\frac{1}{\hbar  v}\int_0^zdz'\,W(z')} e^{i\bk.\br}.
\end{equation}

\section{Spin texture}

To calculate the spin texture we require a microscopic basis of the bulk Dirac equation. We denote this basis as: $\ket{\phi_2^-}$, $\ket{T\phi_2^-}$, $\ket{\phi_1^+}$, $\ket{T\phi_1^+}$, where the pairs of $T$ related states correspond to the two degenerate Kramers sub-spaces of the bulk valence ($\phi_2^-$)  and conduction ($\phi_1^+$) gap edge bands\cite{vol83} ($T$ is time reversal operator). The surface state can then be written as

\begin{eqnarray}
\ket{\Psi} & = & \frac{e^{-i\left(\theta_\bk/2-\pi/4\right)}}{2}
\left(s_+\ket{\phi_1^+} + s_-\ket{\phi_2^-}\right) \nonumber \\
& + &  \frac{l e^{i\left(\theta_\bk/2-\pi/4\right)}}{2}
\left(-s_+\ket{T \phi_1^+} + s_-\ket{T \phi_2^-}\right)
\end{eqnarray}
This has the form
$\ket{\Psi} = \ket{X} + l\ket{TX}$ from which the spin texture $\bmm_\bk = \mel{\Psi}{\bsig}{\Psi}$ is evaluated as

\begin{eqnarray}
m_\bk & =& l\Re\, i e^{-i\theta_\bk}
\Bigg(
\frac{1}{2}\left(1-\frac{\varphi_0}{\Delta_0}\right)\mel{T\phi_2^-}{\bsig}{\phi_2^-} \nonumber \\
& - & 
\frac{1}{2}\left(1+\frac{\varphi_0}{\Delta_0}\right)\mel{T\phi_1^+}{\bsig}{\phi_1^+} \nonumber \\
& + &\sqrt{1-                                                    \frac{\varphi_0^2}{\Delta_0^2}}\mel{T\phi_1^+}{\bsig}{\phi_2^-}
\Bigg).
\label{T}
\end{eqnarray}
This expression shows that by tuning the band bending through the bulk gap $-|\Delta_0| < \varphi_0 < +|\Delta_0|$ the contribution of the conduction, valence, and inter-band contributions to the spin texture are continuously changed by the weight prefactors $\frac{1}{2}\left(1-\frac{\varphi_0}{\Delta_0}\right)$, $\frac{1}{2}\left(1+\frac{\varphi_0}{\Delta_0}\right)$, and $\sqrt{1-                                                    \frac{\varphi_0^2}{\Delta_0^2}}$. Towards the limit of upward band bending, $\varphi_0 \to -|\Delta_0|$, the spin texture is determined fully by the conduction band matrix element, vice versa for the limit of downward band bending the texture is determined fully by the valence band matrix element. The inter-band contribution rises to a maxima at the gap centre, and falls to zero at the gap edges. The possibility therefore exists to tune the surface spin texture via band bending of the bulk spectrum.

\subsection{Spin texture at the $\Gamma$-point Dirac-Weyl cone of SnTe}

\begin{figure}
\centering
\includegraphics[width=0.25\textwidth]{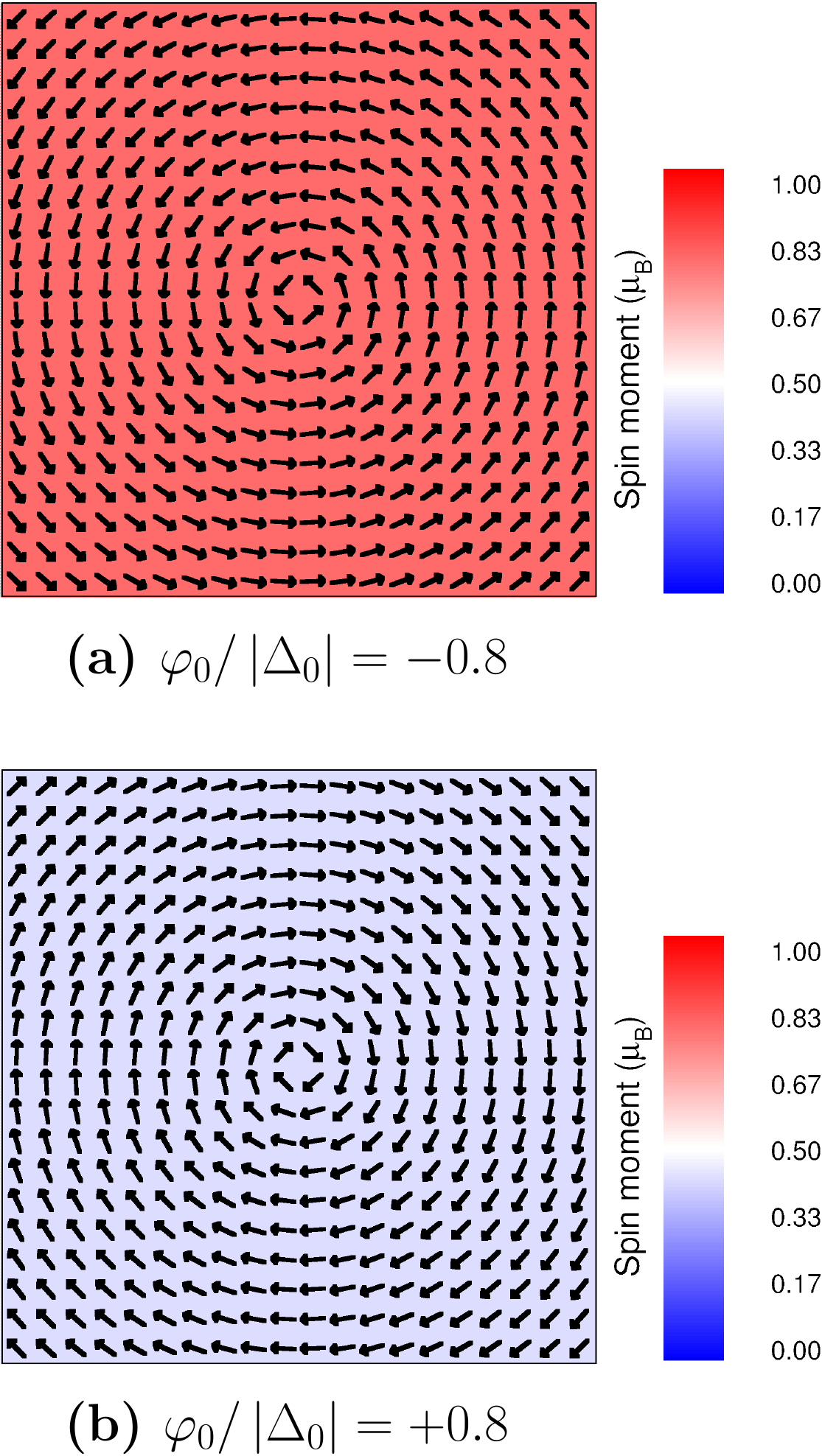}
\caption{Helical textures at the $\Gamma$ point Dirac-Weyl conduction band cone the (111) surface of SnTe for both downward and upward band bending, panels (a) and (b) respectively. Changing the sign of band bending results in a corresponding change in sign of the texture winding number.}
\label{fig2}
\end{figure}

\begin{figure}
\centering
\includegraphics[width=0.35\textwidth]{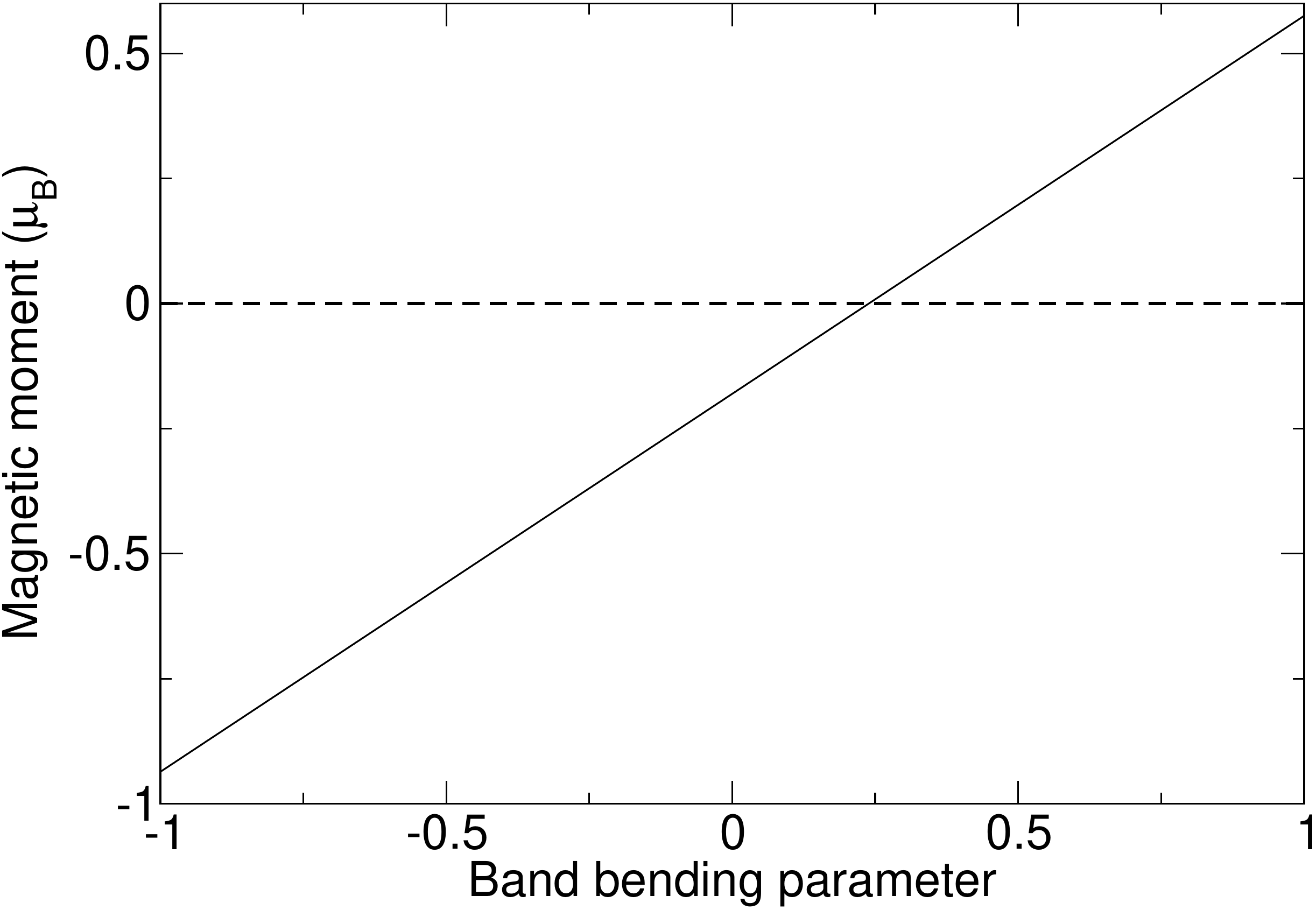}
\caption{Texture magnetic moment for the $\Gamma$ point surface state of the (111) facet of SnTe as a function of band bending. Note that the texture can be completely depolarized by tuning the band bending.}
\label{fig3}
\end{figure}

We now consider the case of the $\Gamma$ point Dirac-Weyl cone a the SnTe (111) surface. As this is separated in energy by $\sim 170$\,meV from the M point cones, a single cone texture is experimentally accessible on this facet\cite{tan13}. From a tight-binding and $\bk.\bp$ analysis\cite{vol78,vol83} one finds for SnTe a bulk Dirac equation with the following basis functions:

\begin{eqnarray}
\ket{\phi_2^-} & = & -\sin\frac{\theta_-}{2}\ket{+\downarrow} + \cos\frac{\theta_-}{2}\ket{0\uparrow} \label{ST1} \\
\ket{T\phi_2^-} & = & -\sin\frac{\theta_-}{2}\ket{-\uparrow} + \cos\frac{\theta_-}{2}\ket{0\downarrow} \\
\ket{\phi_1^+} & = & \cos\frac{\theta_+}{2}\ket{+\downarrow} + \sin\frac{\theta_+}{2}\ket{0\uparrow} \\
\ket{T\phi_1^+} & = & \cos\frac{\theta_+}{2}\ket{-\uparrow} + \sin\frac{\theta_+}{2}\ket{0\downarrow} \label{ST4}
\end{eqnarray}
where $\theta_\pm$ are the spin mixing angles describing the relative strength of the crystal field and spin orbit coupling. For SnTe these take the values of $\theta_+ = 2.63$ and $\theta_- = 1.42$ radians\cite{vol83}. Employing Eq.~\eqref{T} and noting that the interband contribution is zero due to inversion symmetry of the rocksalt lattice we find for the spin texture

\begin{equation}
\bmm_\bk = m\begin{pmatrix} \sin\theta_\bk \\ -\cos\theta_\bk
\\
0\end{pmatrix}
\end{equation}
where

\begin{equation}
m = \frac{l}{2}\left(1-\frac{\varphi_0}{\Delta_0}\right)\cos^2\frac{\theta_-}{2}-\frac{l}{2}\left(1+\frac{\varphi_0}{\Delta_0}\right)\sin^2\frac{\theta_+}{2}
\end{equation}
Thus by tuning the band bending we can vary the texture moment from positive to negative values through a depolarization point at

\begin{equation}
\frac{\varphi_0}{\Delta_0} = \frac{\cos^2\frac{\theta_-}{2} -\sin^2\frac{\theta_+}{2} }{\cos^2\frac{\theta_-}{2} +\sin^2\frac{\theta_+}{2} }
\end{equation}
(see Figs.~\ref{fig2} and \ref{fig3}). The origin of the depolarization phenomena, which should be observable in experiment by appropriate doping of the surface, is that the conduction and valence bands contribute opposite texture winding numbers, see Eq.~\eqref{T}, and so as band bending tunes their relative weight in the surface state a compensation point of depolarization can be reached.
Using the values of $\theta_\pm$, and the rather large band bending that appears to be the case for the Sn terminated facet at which the Dirac-Weyl cone is observed in experiment, we find texture winding number in agreement with {\it ab-intio} calculation\cite{shi14,saf13}, but with a moment somewhat reduced (we find $\sim 0.5\mu_B$ with {\it ab-initio} finding $\sim 0.7\mu_B$)\cite{shi14}.

\subsection{The spin texture more generally}

\begin{figure*}
\centering
\includegraphics[width=0.95\textwidth]{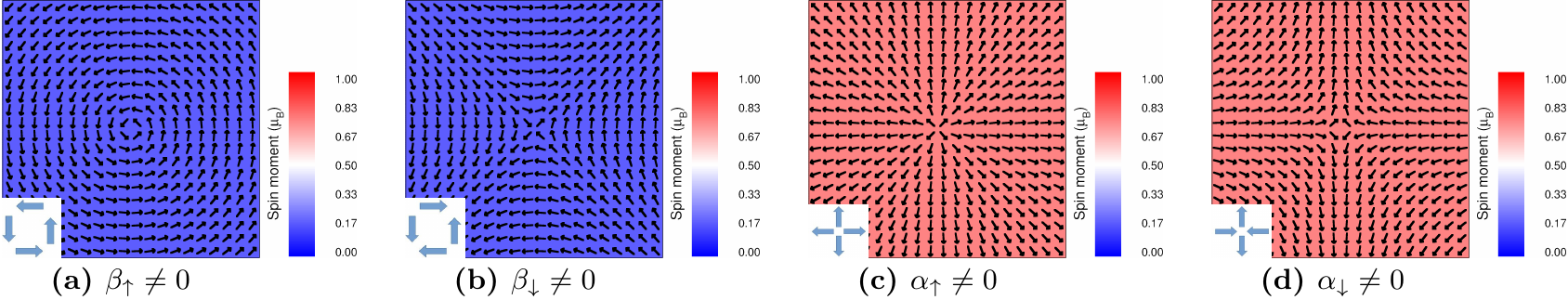}
\caption{The four spin textures possible for general bulk gap edge states: (a) helical, (b) hyperbolic, (c) hedgehog, and (d) hyperbolic hedgehog, with magnitude of the spin moment is indicated by the colour. These correspond to the cases in which only one of the $\alpha_\sigma$ and $\beta_\sigma$ parameters of Eq.~\eqref{GST} is allowed to be non-zero, as indicated in the figure caption. Note that these parameters in turn depend on the bulk microscopic band edge wavefunctions, see Eqs.~\eqref{a}-\eqref{b}. In all plots the surface band bending is set to zero and the texture is shown for the conduction band cone. The inset figure indicates the basis structure of the texture in each case.}
\label{fig4}
\end{figure*}

We now consider what forms of spin texture are in principle possible within the supersymmetric theory of a topological insulator. We therefore take a general form of a gap edge wavefunction

\begin{equation}
\ket{\phi_i} = \sum_{m\sigma} d_{m\sigma}^{(i)} \ket{m\sigma}
\end{equation}
where $\ket{m\sigma}:= \ket{m}\otimes\ket{\sigma}$ is a spin-angular momentum function. At this stage we will make no assumptions concerning the angular momentum character of the gap edge wavefunction. The $T$ conjugated partner of this wavefunction is

\begin{equation}
\ket{T \phi_i} = \sum_{m\sigma} (-1)^m \mathrm{sign}(\sigma) {d_{m\sigma}^{(i)}}^\ast \ket{-m-\sigma}
\end{equation}
From $\ket{\phi_i}$ and $\ket{T \phi_i}$ we can now obtain a general form for the matrix elements involved in the spin texture formula Eq.~\eqref{T}

\begin{equation}
\mel{T \phi_j}{\bsig}{\phi_i} = 
(a_{ji}^\sigma+ib_{ji}^\sigma)(\sigma,i,0) + (c_{ji}^\sigma+id_{ji}^\sigma)(0,0,1)
\label{M}
\end{equation}
where
\begin{eqnarray}
a_{ji}^\sigma = \sum_m (-1)^m \Re\,d_{-m\sigma}^{(j)} d_{m\sigma}^{(i)} \label{start} \\
b_{ji}^\sigma = \sum_m (-1)^m \Im\,d_{-m\sigma}^{(j)} d_{m\sigma}^{(i)} \label{34} \\
c_{ji}^\sigma = \sum_m (-1)^m \Re\,d_{-m-\sigma}^{(j)} d_{m\sigma}^{(i)} \label{35} \\
d_{ji}^\sigma = \sum_m (-1)^m \Im\,d_{-m-\sigma}^{(j)} d_{m\sigma}^{(i)} \label{36}
\end{eqnarray}
Evaluating the spin texture, Eq.~\ref{T}, employing general forms for the conduction $\ket{\phi_1^+}$ and valence $\ket{\phi_2^-}$ states given by Eq.~\eqref{M} leads to the result

\begin{equation}
\bmm_\bk = l \sum_\sigma
\begin{pmatrix}
\sigma(\alpha_\sigma \cos\theta_\bk + \beta_\sigma \sin\theta_\bk)\\\alpha_\sigma\sin\theta_\bk-\beta_\sigma\cos\theta_\bk\\
-\gamma_\sigma\sin\theta_\bk - \delta_\sigma\cos\theta_\bk
\end{pmatrix}
\label{GST}
\end{equation}
where

\begin{eqnarray}
\alpha_\sigma & = & \frac{1}{2}\left(1+\frac{\varphi_0}{\Delta_0}\right)b_{11}^\sigma-\frac{1}{2}\left(1-\frac{\varphi_0}{\Delta_0}\right)b_{22}^\sigma\nonumber\\
&&-\sqrt{1-                                                    \frac{\varphi_0^2}{\Delta_0^2}}a_{12}^\sigma \label{a} \\
\beta_\sigma & = & -\frac{1}{2}\left(1+\frac{\varphi_0}{\Delta_0}\right)a_{11}^\sigma+\frac{1}{2}\left(1-\frac{\varphi_0}{\Delta_0}\right)a_{22}^\sigma\nonumber\\
&&-\sqrt{1-                                                    \frac{\varphi_0^2}{\Delta_0^2}}b_{12}^\sigma \label{b}
\end{eqnarray}
with similar formula for  $\delta_\sigma$ and $\gamma_\sigma$ obtained by $a\to c$ and $b\to d$ in Eqs.~\eqref{a} and \eqref{b} respectively. 

Equations \eqref{start}-\eqref{b} represent the most general form of the surface spin texture possible for a Dirac-like bulk spectrum.
Before analyzing this result, however, two caveats are in order: (i) We are considering the case of single Dirac-Weyl topological surface state, a special circumstance occurring either in "strong" topological insulators with a single $\Gamma$-point surface state, or resulting from energy shifts among the several surface states of a "weak" topological insulator, as happens on the (111) facet of SnTe where $\Gamma$ cone is energetically isolated from the three M point cones. More complex spin textures can arise from the coupling of distinct surface states\cite{liu13}, a situation we do not consider here. (ii) A continuum Hamiltonian and its microscopic basis must be obtained as a low energy approximation to a corresponding tight-binding description of the system of interest, and by introducing general bulk gap edge states this link in broken. On the other hand, it should be stressed that the Dirac equation is a generic low energy description of Kramer's degenerate conduction and valence bands, with wide applicability. Moreover our aim here is to describe formally the range of texture types possible within the supersymmetric theory of a TI surface state arsing from a Dirac-like bulk spectrum, rather than provide specific examples.

\subsection{Four spin textures}

We will keep the band bending fixed at zero, and so the distinct types of spin textures that we will explore in this section are due only to changes in the microscopic gap edge states of the bulk.

We first consider the out-of-plane polarization of the spin texture.
From Eqs.~\eqref{35} and \eqref{36} we see that this can arise only if the gap edge states contain a finite contribution from both $(+m,+\sigma)$ and $(-m,-\sigma)$ angular momentum spin functions. For the gap edge states of SnTe this is not the case, see Eqs.~\eqref{ST1}-\eqref{ST4}. However, any perturbation of SnTe that would lift parity symmetry of the bulk gap states and allow inter-band matrix elements would automatically generate an out-of-plane polarization. Interestingly, a finite contribution from both $(+m,+\sigma)$ and $(-m,-\sigma)$ functions can evidently also be achieved by rotation within the degenerate Kramers pairs of each band, which may be induced by misalignment between the surface normal and the bulk coordinate system. In fact in \emph{ab-initio} calculations\cite{shi14} of the (111) facet M point texture (which is misaligned by the angle $\cos^{-1}\frac{1}{3}$) an out-of-plane polarization is seen.

Turning to the in-plane spin polarization one notes that despite the general form of the bulk wavefunction the in-plane spin texture depends on only four coefficients, $\alpha_\uparrow$, $\alpha_\downarrow$, $\beta_\uparrow$, and $\beta_\downarrow$, and so only four texture types can be achieved. These are shown in Fig.~\ref{fig4} and we now describe the origin of each in turn.

The case of SnTe considered in the previous section corresponds to 
$\beta_\uparrow\ne0$ with all other coefficients zero, generating the familiar helical texture, Fig.~\ref{fig4}a. Flipping the direction of the spin in the basis wavefunctions to give conduction and valence states 

\begin{eqnarray}
\ket{\phi_2^-} & = & -\sin\frac{\theta_-}{2}\ket{+\uparrow} + \cos\frac{\theta_-}{2}\ket{0\downarrow} \\
\ket{\phi_1^+} & = & \cos\frac{\theta_+}{2}\ket{+\uparrow} + \sin\frac{\theta_+}{2}\ket{0\downarrow}
\end{eqnarray}
result in $\beta_\downarrow\ne0$ with all other coefficients zero. This generates the hyperbolic texture shown in Fig.~\ref{fig4}b. It is remarkable that the texture depends profoundly on the alignment of the bulk band edge spin and the surface normal. A possible way in which such a texture could be realized would again be misalignment between the surface normal and the bulk coordinate system, and in fact for the (111) facet of SnTe a hyperbolic M point texture can be obtained.

Finally we have the cases where $\alpha_\uparrow\ne0$ or
$\alpha_\downarrow\ne0$ with all other coefficients zero. The simplest orthogonal $p$-band bulk band edge states that can generate $\alpha_\uparrow\ne0$ are

\begin{eqnarray}
\ket{\phi_2^-} & = & i\sin\frac{\theta_-}{2}\ket{-\uparrow} + \cos\frac{\theta_-}{2}\ket{+\uparrow} \\
\ket{\phi_1^+} & = & -i\cos\frac{\theta_+}{2}\ket{-\uparrow} + \sin\frac{\theta_+}{2}\ket{+\uparrow}
\end{eqnarray}
As shown in Fig.~\ref{fig4}c this generates a "hedgehog" texture, and flipping of the spin direction then generates the corresponding "hyperbolic hedgehog", Fig.~\ref{fig4}d.

In contrast to the previous two cases these can not be achieved with bulk gap edge states that are eigenfunctions of the $J_z$ operator; an imaginary result for the product of coefficients of angular momentum spin functions ($+m,+\sigma)$ and $(-m,+\sigma)$ evidently requires that $m\ne0$ (see Eq.~\eqref{34}), and such a state cannot be an eigenfunction of $J_z$. A go around for this would be via interband contributions (the $a_{12}^\sigma$ term in Eq.~\eqref{a}, however such interband contributions are generally forbidden by symmetry.

Finally, we note that spin texture depolarization seen for the $\Gamma$ point Dirac-Weyl cone on (111) facet of SnTe may still occur for the more general gap edge states described here. Ignoring interband contributions gives a compensation point of $\varphi_0/\Delta_0 = -(a_{11}^\sigma-a^\sigma_{22})/(a_{11}^\sigma-a^\sigma_{22})$ and when $\mathrm{sign}(a_{11}^\sigma a_{22}^\sigma) > 0$ this occurs for band bending within the bulk gap, as $\sqrt{1-                                                    \frac{\varphi_0^2}{\Delta_0^2}}$ at the limits $\varphi_0/\Delta_0=\pm1$ interband contributions to the texture will only shift the depolarization band bending. For $\mathrm{sign}(a_{11}^\sigma a_{22}^\sigma) > 0$ spin texture depolarization is not guaranteed to occur within the gap, but may do so depending on relative strength of intra- and inter-band contribution to the spin texture.

\section{Discussion}

We have applied the supersymmetry approach to the material-vacuum junction of a topological insulator with the material side modeled by a Dirac equation. Our two main conclusions are: (i) That the spin texture of the TI state can be qualitatively changed by the physics of the bulk band edge states. For a Dirac type bulk spectrum we find that 4 distinct spin texture types are supported by the interface state, and which of these is realized depends solely on the bulk gap edge states. (ii) That band bending determines the relative contributions of the conduction, valence, and conduction-valence bulk matrix elements in the spin texture; for zero band bending these enter the texture with equal weight, while at the limits of upwards (or downwards) band bending the texture completely dominated by the conduction (or valence) band of the bulk. 

For the case the SnTe (111) $\Gamma$-point Dirac cone, sweeping the band bending tunes the helical texture from positive to negative winding number through a point of complete depolarization. The existence of a depolarization point is a general feature, however it is not guaranteed to be accessible without collapsing of the surface state by tuning the band bending outside the bulk gap. Utilizing tight binding derived SnTe bulk gap edge wavefunctions taken from, we find agreement with {\it ab-initio} calculations for the spin texture, although with the value of the spin moment somewhat reduced.

Replacing the SnTe wavefunctions derived from tight-binding by a general form of the gap edge states $\ket{\phi_i} = \sum_{\sigma m} d_{m \sigma} \ket{m\sigma}$ find two new spin textures are in principle possible: hyperbolic and hedgehog. The former can be accessed by rotations with the Kramers degenerate sub-space of the bulk edge bands (e.g. by misorientation of the bulk Dirac equation from the surface), however the latter demands low symmetry not found in TIs but which could perhaps be obtained by either mechanical or growth induced strain.

%\bibliographystyle{unsrt}
%\bibliography{literature_topo}

\end{document}